\documentclass[prl,twocolumn,showpacs,preprintnumbers,amsmath,amssymb]{revtex4}
\usepackage{txfonts}
\begin{document}
\title{Entanglement between living bacteria and quantized light witnessed by Rabi splitting}
\author{C. Marletto, D. M. Coles, T. Farrow, V. Vedral}
\affiliation{Clarendon Laboratory, University of Oxford, Parks Road, Oxford OX1 3PU and \\
Department of Physics and Astronomy, University of Sheﬃeld, Sheﬃeld S3 7RH, UK5 \\
Centre for Quantum Technologies, National University of Singapore, 3 Science Drive 2, Singapore 117543 and\\
Department of Physics, National University of Singapore, 2 Science Drive 3, Singapore 117542}
\begin{abstract}
We model recent experiments on living sulphur bacteria interacting with quantised light, using the Dicke model. The strong coupling achieved between the bacteria and the light indicates that during the experiment the bacteria (treated as dipoles) and the quantized light are entangled. The vacuum Rabi splitting, which was measured in the experiment for a range of different parameters, can be used as an entanglement witness. 
\bigskip
\end{abstract}
\maketitle
Witnessing quantum effects in living systems was long considered an impossible task, even by the pioneers of quantum theory, such as Bohr~\cite{BOHR}. Recent advances in theoretical and experimental techniques are bringing us closer to accomplishing it. Entanglement has extensively been investigated, and even detected, in various many-body systems \cite{Amico}. Since living systems are special cases of many-body systems, they can be analysed with the same methods. In this paper we focus on experiments where living sulphur bacteria are entangled with a quantised field of light~\cite{Coles}. This is particularly exciting, since the quantised nature of light was at the heart of the complementarity that Bohr thought would ultimately make it impossible for us to detect quantum effects in a living entity.
Green sulphur bacteria are found in anaerobic environments rich in sulphur compounds, such as microbial mats and around hot springs~\cite{Prokaryotes}. While they are photosynthetic, they can survive in locations where light intensity drops to only a few hundred photons per second per bacteria\cite{Overmann}. The bacteria have evolved large antenna complexes called chlorosomes, which are large aggregates of approximately 200,000 self-assembled bacteriochlorophyll (BChl) molecules. Each bacterium contains around 200 chlorosomes, where the dense packing of the molecules and their high dipole moments result in high oscillator strengths that make the bacteria (and organic matter in general) good candidates for strong coupling. Their size, approximately 2 $\muup$m $\times$ 500 nm, means that they can also be inserted into a micron-sized optical microcavities with well-defined photon mode energies. The strong coupling condition is met when the leakage of the light trapped in the microcavity is slow compared to the energy exchange rate between the light and bacteria.

We resort to a quantum-mechanical model where the BChl molecules situated in living bacteria are modeled as electric dipoles and reduced to two-level systems. Each of these dipoles is strongly coupled to the single frequency of light in the cavity through the Jaynes-Cummings Hamiltonian. 
The bacteria are modeled as a system of $N$ two-level atoms, collectively interacting with the field of a single-mode cavity whose annihilation (creation) operator is $a^{\dagger}$ ($a$). 
Each two-level atom is modeled as a pseudo-spin whose Pauli spin matrices are
$\{\sigma^i_{\pm}, \sigma^i_z\}_{(i=1, .., N)}$. We assume a realistic small-sized bacterial sample, neglect the variations of the cavity field at its location and take the coupling strength as uniform. Then, we introduce the total angular momentum $J$ of the atomic sample with components $J_{\pm} =\sum \sigma^i_{\pm}$ and $J^z =\sum \sigma^i_z$ and consider time-independent single-spin energy splittings $\omega_b$. The Hamiltonian of the system in the dipole approximation thus reads (with $\hbar = 1$ throughout)
\[
H_0 = \omega_a a^{\dagger}a + \omega_b J^{z} + g (a^{\dagger} + a)(J_+ + J_{-}), 
\]
where $\omega_a$ is the frequency of the cavity and $g$ is the vacuum Rabi frequency (ignoring the vacuum contribution without any loss of generality and assuming $\hbar=1$). The parameter $j$ is the cooperation number in the Dicke theory~\cite{dicke}, that is an eigenvalue of $J^2$ which, together with the eigenstates of $J_z$, is used to build the Dicke states. The ensemble of $N$ two-level atoms is then described as a pseudo-spin of size $j = N/2$. 
We can now apply the Holstein-Primakoff transformation, defined as
\[
{\displaystyle J_{+}=\hbar {\sqrt {2j}}{\sqrt {1-{\frac {b^{\dagger }b}{2j}}}}\,b~,\qquad J_{-}=\hbar {\sqrt {2j}}b^{\dagger }\,{\sqrt {1-{\frac {b^{\dagger }b}{2j}}}}~,
}
\]
and
\[
\qquad J_{z}=\hbar (j-b^{\dagger }b)~.
\]
In the limit $j>>1$ (thermodynamic limit) we obtain
\[
H_0 \approx \omega_a a^{\dagger}a + \omega_b b^{\dagger}b + \sqrt{N} g (a^{\dagger} + a)(b^{\dagger} + b), 
\]
The eigenstates of this Hamiltonian are typically entangled states between a number of excitations in one of the oscillators (representing excitons in the bacteria) and the other harmonic oscillator (representing photons in a single mode of light). Such ``dressed" states between excitons and photons are known as polaritons. These quasi-particles have a mixed physical character that is part-matter and part-light. 

The dense packing and high oscillator stengths in organic media make them well suited to giving rise rise to these quasi-particles via strong coupling with light. A special class of these particles, known as organic surface plasmon polaritons, have been showcased in a series of recent experiments where quantized light and organic matter were strongly coupled via cavities and organic semiconductors~\cite{lid, barbie}, down to the level of manipulating single organic molecules at room temperature~\cite{chik}. Besides the inherent interest that the hybrid states of light and matter present, they are predicted to find uses such as in controlling the chemical kinetics in a wide class of photochemical reactions~\cite{gal}.

To demonstrate entanglement we use entanglement witnesses, which are physical observables that ``react'' differently to entangled and disentangled states. Here we will only focus on ground-state entanglement since the relevant frequencies are at least one order of magnitude larger than $kT$ at room temperature at which the experiment was carried out. 
In our case of two effective coupled harmonic oscillators we can use the energy as a witness. For simplicity, we assume that $\omega_a=\omega_b=\omega$, i.e. that the exciton frequency is on-resonance with the light in the cavity. The energy of separable states of the form $\sum_{n,m} p_{n,m} |nm\rangle\langle nm|$ can never be smaller than
\[
\langle H_0\rangle_{sep} \geq 0
\]
In the real ground state the energy is lower by the vacuum Rabi splitting $\sqrt{N}g$ (we are ignoring the vacuum contribution without loss of generality). Hence the ground state is an entangled state. We can therefore conclude from the measurements of Rabi splitting on two coupled harmonic oscillators assumed for light and bacteria (treated as a large collection of dipoles), that they are entangled. Specifically, the entangled subsystems are the excitons in the bacteria and the photons in the cavity. 
The amazing fact is that the bacteria can actually stay alive during the experiments, despite the fact that they are strongly coupled to the cavity light. (The test of being alive in \cite{Coles} was a form of homeostasis - the fact that the bacteria repelled a certain dye molecule during the experiment, which they absorb when dead). 
The entanglement witness above is also effectively a measure of entanglement. Namely, the ground state is entangled so long as $g>0$. The amount of entanglement between bacteria and light could be quantified by the linear entropy of either of the subsystems. The linear entropy is defined as the complement of the purity of the state, i.e., $1-{\rm Tr}\{\rho^2\}$ where $\rho$ is the reduced density operator of either of the subsystem (obtained by tracing out the degrees of freedom of the other subsystem). A simple calculation shows that the linear entropy is $\left(\frac{\sqrt{N}g}{\omega}\right)^2$, where $\sqrt{N}g < \omega$. 
In \cite{Coles} the highest coupling achieved was $\sqrt{N}g \approx 0.2 \omega$ and therefore the amount of entanglement measured was $0.04$. 
A question could be raised as to whether there might be other processes that could affect this entanglement. For instance, the cavity could leak photons which could spoil the coherence of polaritons. However, we take it as an observational fact that polaritons last long enough to be detected and must therefore preserve coherence during this period. Excitons could also dissipate within bacteria through interactions with phonons, for instance, but these processes are typically on much longer timescales than the cavity photon lifetime. 
The witness of entanglement between light and bacteria outlined here has the property of being one of the easiest to access experimentally. However, it is worth pointing out that the same results (namely the observation of the vacuum Rabi splitting) can equally well be explained by a completely classical analysis. There is no contradiction with the entanglement witness in our model, because the witness already assumes that both systems are quantum and checks whether the subsystems are entangled. In the classical analysis, on the other hand, the cavity mode is represented by a simple harmonic oscillator and is coupled to $N$ classical dipoles each of which represents a BChl molecule within the bacteria. The emission spectrum obtained via the multi-beam interference technique (assuming that the cavity mode is on resonance with the dipole frequency) \cite{Zhu} exhibits a splitting in the intensity peaks separated by the classical analogue of the vacuum Rabi splitting, given by:
\begin{equation}
\Omega = d \sqrt{\frac{N \omega}{\epsilon_0 L_c}}
\end{equation}
where $\epsilon_0$ is the vacuum permittivity, $L_c$ is the length of the cavity, and $d$ is the dipole moment. This classically obtained result is exactly the same as that obtained by the full quantum analysis above ($\sqrt{N} g$ where $g$ is the vacuum Rabi frequency) and, interestingly, this also includes the square root scaling with the number of oscillators (i.e. the molecules). 

One important subtlety should be mentioned here - relating to the indirect inference of entanglement in our system. If instead of the fully classical model, or indeed the fully quantum model, we use a semi-classical model (where light is treated as a quantum system because we have direct experimental evidence of this fact), we can no longer obtain the vacuum Rabi splitting. This is a special case of a more general feature where semi-classical models can never fully reproduce the quantum features. One could therefore argue that since light is known to be quantum and the Rabi splitting is observed, the bacteria (more precisely, whatever degree of freedom within bacteria couples to light) also have to be quantum. The fact that the semi-classical model fails to reproduce certain correlations between two subsystems is not unusual; in fact, it also features in cryptographic protocols, \cite{ACIN}, where four bits can reproduce some correlations of a singlet, but one qubit and two bits cannot.

There are however more refined and direct levels of probing and confirming entanglement between two subsystems, here being the two harmonic oscillators, one representing the bacteria and the other the cavity mode. At the most detailed level we have Bell's inequalities \cite{Bell}. To detect a violation of the inequality, one would need to probe the bacteria and light separately, by measuring two suitably chosen conjugate observables for each. For instance, one could measure (effective) $x$ and $p$ on both the dipoles in the bacteria and the cavity light. This is of course much harder to achieve experimentally. Harder still would be to spatially separate the bacteria and light to ensure that there is no signal propagation between the two while measurements are carried out (to ensure that the locality loophole is closed). 
Another possible line of probing the quantum character of the bacteria-light interaction is to investigate the quantum coherent dynamics of polaritons. A less challenging experiment would be the observation of Rabi flopping between the two Rabi split levels associated with the polariton. However, given that the spectrum obtained in \cite{Coles} is effectively just the modulus squared of a Fourier Transform of the Rabi flopping, it is still questionable whether this would be a conclusive proof of quantum coherence, while it would certainly not be a witness of entanglement since the bacteria and light are treated as one quantum system, namely a polariton. At the harder end of the spectrum would be violating the Leggett-Garg inequality and similar (e.g. \cite{Legg}). This would again require identifying suitable observables to be measured on the polariton, as well as devising a means of performing a non-invasive (weak) measurement.
\\\textit{Acknowledgments}: VV and TF thank the Oxford Martin School, Wolfson College and the University of Oxford, the Leverhulme Trust (UK), the John Templeton Foundation, the EU Collaborative Project TherMiQ (Grant Agreement 618074), the COST Action MP1209, the EPSRC (UK) and the Ministry of Manpower (Singapore). This research is also supported by the National Research Foundation, Prime Minister’s Office, Singapore, under its Competitive Research Programme (CRP Award No. NRF- CRP14-2014-02) and administered by Centre for Quantum Technologies, National University of Singapore.

\end{document}